\documentclass[cit]{PoS}

\usepackage{graphicx}
\usepackage{graphics}

\def \vhel{\ifmmode{~V_{{\rm HEL}}}\else{~$V_{{\rm HEL}}$}\fi}
\def \vsys{\ifmmode{~V_{{\rm SYS}}}\else{~$V_{{\rm SYS}}$}\fi}
\def \HA {\ifmmode{{\rm\H}\alpha}\else{${\rm\ H}\alpha$}\fi}

\def \msun{\ifmmode{{\rm\ M}_\odot}\else{${\rm\ M}_\odot$}\fi}

\def \myr{\ifmmode{{\rm\ M}_\odot{\rm\ yr}^{-1}}
         \else{${\rm\ M}_\odot$ yr$^{-1}$}\fi}

\def \mdot{\ifmmode{\dot{M}}\else{$\dot{M}$}\fi}
\def \tena#1 #2 {\ifmmode{#1 \times 10^{#2}}\else{$#1 \times 10^{#2}$}\fi}
\def \kms{\ifmmode{~{\rm km\,s}^{-1}}\else{~km s$^{-1}$}\fi}

\def \apj{ApJ}
\def \mnras{MNRAS}

\def \aap{A\&A}

\def \apjl{ApJL}
\def \nat{Nat}

\title{VLBI constraints on the ``jet-line'' of Cygnus X-1}

\ShortTitle{The ``jet-line'' of Cygnus X-1}

\author{\speaker{Anthony Rushton},$^{ab}$ James Miller-Jones,$^{c}$ Zsolt Paragi,$^d$ Thomas Maccarone,$^e$ Guy Pooley,$^f$ Valeriu Tudose,$^d$ Rob Fender,$^e$ Ralph Spencer,$^g$ Vivek Dhawan,$^h$
and Michael Garrett$^{i}$\\
\llap{$^a$}European Southern Observatory\\
Karl-Schwarzschild-Str 2, 85748 Garching, Germany\\
\llap{$^b$}Onsala Space Observatory\\
SE-439 92, Sweden\\
E-mail: \email{Anthony.Rushton at eso.org}\\
\llap{$^c$}ICRAR - Curtin University of Technology\\
 GPO Box U1987, Perth, WA 6845, Australia\\
\llap{$^d$}Joint Institute for VLBI in Europe\\
Postbus 2, 7990 AA Dwingeloo, The Netherlands\\
\llap{$^e$}School of Physics and Astronomy, University of Southampton\\
Highfield, Southampton SO17 1BJ, UK\\
\llap{$^f$} Cavendish Laboratory\\
J. J. Thomson Avenue, Cambridge CB3 0HE, UK\\
\llap{$^g$}Jodrell Bank Centre for Astrophysics, School of Physics and Astronomy, University of Manchester\\
M13 9PL, UK\\
\llap{$^h$}NRAO Domenici Science Operations Center\\
1003 Lopezville Road, Socorro, NM 87801, USA\\
\llap{$^i$}ASTRON\\
Oude Hoogeveensedijk 4, 7991 PD Dwingeloo, The Netherlands\\}

\abstract{Results are presented from recent VLBI observations of Cygnus X-1 during X-ray spectral state changes. Using the EVN in e-VLBI mode and the VLBA with disk recording, we observed the X-ray binary at very high angular resolution and studied changes in the compact jets as the source made transitions from hard X-ray states to softer states.  The radio light curves show that these transitions were accompanied by radio flaring events followed by a quenching of the radio emission, as expected from the current paradigm for disc-jet coupling in X-ray binaries.  While we see structural changes in the compact jets during these transitions, there was no evidence for the expected ejection of bright, relativistically-moving jet knots.  However, we find strong evidence that the jet does not switch off completely in the soft X-ray state of Cygnus X-1, such that a weak, compact jet persists during this phase of radio quenching.}

\FullConference{10th European VLBI Network Symposium and EVN Users Meeting: VLBI and the new generation of radio arrays\\
		September 20-24, 2010\\
		Manchester UK}

\begin{document}

\section{Introduction}

Cygnus X-1 (a.k.a.\ HDE 226868, V1357 Cygni) was first discovered in 1964 during one of two Aerobee X-ray surveys~\cite{1965Sci...147..394B} . Radio emission was later discovered with the Green Bank Interferometer~\cite{1971ApJ...168L..21H} and also by using the Westerbork Synthesis Radio Telescope~\cite{1971Natur.232Q.246B}. Shortly following this came the discovery of an optical counterpart as a $9^{\rm{th}}$ magnitude O9.7 Iab star with an orbital period of 5.6~days~\cite{1987SvA....31..170K}. Spectral line observations have shown the source to comprise a binary system with a supergiant companion (with a mass between 20 and 33~\msun); hence it is a high-mass X-ray binary (XRB), in which the accretor is a compact object with a mass between 7 and 16~\msun~\cite{1986ApJ...304..389G}. Cygnus~X-1 is therefore one of the best Galactic black hole candidates and has been well studied across the electromagnetic spectrum.

The first resolved high-resolution Very Long Baseline Interferometry (VLBI) observations of Cygnus~X-1 were three closely spaced joint Very Long Baseline Array (VLBA) and Very Large Array (VLA) observations at 8.4~GHz taken in 1998 August~\cite{2001MNRAS.327.1273S}. The results showed an extended $\sim15$~mas-scale jet to the north-west of an bright core. This demonstrated the presence of a resolved compact steady jet whilst the source was in a low/hard state. Since flat-spectrum radio emission is always associated with the low/hard state of Cygnus X-1, it is suggested that this resolved steady jet may always be present in this state. Furthermore, a large-scale `lobe' structure~\cite{2005Natur.436..819G} is aligned with the position angle of the VLBI-scale jet, suggesting that the jets have inflated a bubble in the surrounding medium and thereby constraining the time-averaged power of the jets.  The only detection to date of a discrete, transient jet from Cygnus X-1 appears to have been associated with an X-ray spectral state change, as the source crossed the so-called ``jet-line'' at a hardness ratio of $\sim0.3$ in an X-ray hardness intensity diagram (HID)~\cite{2006MNRAS.369..603F}.
 
\section{Observations}

Previous VLBI observations of Cygnus X-1 have only been taken during the predominant low-hard state~\cite{2001MNRAS.327.1273S}, in which the source can remain for many years. Our goal was to observe the XRB with high-resolution VLBI during a hard-to-soft X-ray spectral state change.  To identify a state change, we relied on quasi-daily radio and X-ray monitoring data from the Arcminute Microkelvin Imager (AMI)  telescope in Cambridge at 15 GHz and the All Sky Monitor (ASM) on board the {\it Rossi X-ray Timing Explorer (RXTE)} between 2 and 12~keV, and also on reports in \textit{The Astronomer's Telegram}\footnote{http://www.astronomerstelegram.org}(ATel).

During 2009 June, Cygnus X-1 started to show evidence of an X-ray spectral state change, from hard to soft, for the first time in 3--5 years. We therefore submitted target-of-opportunity (ToO) requests to the European VLBI Network (EVN) and VLBA which were both approved. The first VLBI observations were taken on 2009~June~28~(MJD~55\,010) using the VLBA at 4~cm with a recording rate of 256~Mbps, corresponding to a bandwidth of 32 MHz per polarisation, with dual polarisation. Shortly after the observations however, the 2--12\,keV X-ray flux started to decay and the source quickly returned to the low-hard state, suggesting the event was a ``failed'' state change (e.g.~\cite{2003A&A...407.1039P}). We therefore cancelled the rest of the approved VLBI runs and waited to trigger on another state change.

It was not until 2010 July that Cygnus X-1 showed any further signs of an X-ray spectral state change. Following a series of ATel reports suggesting a state transition starting around 2010~June~28~\cite{2010ATel.2711....1N,2010ATel.2714....1R,2010ATel.2715....1S}, it became apparent that a full state change was occurring. Once again we requested (and had approved) ToO observations with the VLBA and EVN. Two initial observations were taken with the EVN on 2010 July 8 and 10 in near real-time e-VLBI mode.  The telescopes participating were Jodrell Bank MkII, Knockin, Cambridge, Westerbork, Effelsberg, Torun, Yebes, Medicina, Onsala 25-m and Sheshan. Data were transferred from each antenna to the correlator using high-speed dedicated network links; connection rates of up to 1024~Mbps were sustained per antenna, yielding maximum bandwidths of up to 128~MHz per polarisation, with dual polarisation. The rapid e-VLBI imaging of the source was then used to help schedule recorded VLBA observations.  From the unresolved 5-GHz e-EVN image, we suspected that any transient ejecta might have expanded sufficiently to be resolved out, so were motivated to schedule the VLBA observations in the simultaneous 4/13~cm mode, using the lower frequency to probe larger spatial scales.  A total of five VLBA  epochs were taken, on July 12, 15, 17, 19 and 22 using all ten antennas, with a recording rate of 512 Mbps (divided equally between S and X bands), give a total bandwidth for each frequency band of 32~MHz per polarisation, with dual polarisation.

All VLBI observations were phase referenced to the calibrator source J1953+3537. In the last four VLBA observations we inserted a geodetic VLBI calibrator block, in order to improve the astrometric accuracy and image quality of the target source without using self-calibration (see \textsc{aips}  Memo 110 for more details). All data were then reduced using the standard \textsc{aips} VLBI algorithms (i.e. \textsc{vlbautil}), using the standard EVN pipeline for initial processing of the e-EVN data.  Table~\ref{table:journal} gives a summary of all VLBI observations.

\begin{table*}[h]
\footnotesize
\begin{center}\begin{tabular}{|l|ccccccc|}
\hline
Epoch & Start date & End date  & $\lambda$ & Array  & $I_{\rm{total}}$   & Noise & Resolved \\
(MJD) &  (UT) &  (UT)       &   (cm) & &   (mJy)          & ($\mu$Jy/bm)    &   ?   \\
\hline 
55\,010 & 09-Jun-28 06:00:00 & 09-Jun-28 14:29:58 & 4 & VLBA &  $16.8\pm3.8$ &  $45$ & Yes \\
55\,026 & 09-Jul-14 02:14:09 & 09-Jul-14 10:07:49 & 4 / 20 & VLBA &  $18.8\pm1.1/12.0\pm1.3$ &  $123/155$&  Yes \\
55\,386 & 10-Jul-08 18:31:30 & 10-Jul-09 05:39:10 & 6 & e-EVN &  $15.2\pm0.1$ &  $116$& No \\
55\,388 & 10-Jul-10 18:16:30 & 10-Jul-11 05:24:10 & 6 & e-EVN &  $5.0\pm0.1$ &  $97$&   No \\
55\,389 & 10-Jul-12 08:44:59 & 10-Jul-12 11:15:00 & 4 / 13 & VLBA &  $5.9\pm0.2/5.3\pm0.2$ &  $155/184$&  No \\
55\,392 & 10-Jul-15 06:27:51 & 10-Jul-15 11:33:29 & 4 / 13 & VLBA &  $3.8\pm0.1/5.2\pm0.2$ &  $104/157$&  No \\
55\,394 & 10-Jul-17 06:23:26 & 10-Jul-17 11:33:09 & 4 / 13 & VLBA &  $3.1\pm0.1/3.1\pm0.2$ &  $102/172$&   No \\
55\,396 & 10-Jul-19 06:27:34 & 10-Jul-19 11:33:59 & 4 / 13 & VLBA &  $1.4\pm0.1/1.2\pm0.2$ &  $95/159$&  No \\
55\,399 & 10-Jul-22 06:25:17 & 10-Jul-22 11:32:40 & 4 / 13 & VLBA &  $2.7\pm0.1/5.7\pm0.2$ &  $99/206$&   No \\
\hline
\end{tabular}\end{center}
\caption{\label{table:journal}Details of the VLBI observations of Cygnus X-1 taken in 2009 and 2010.}
\end{table*}

\vspace{-0.45cm} 

\section{Results}

\subsection*{Observations of summer 2009}

\begin{figure}[h]
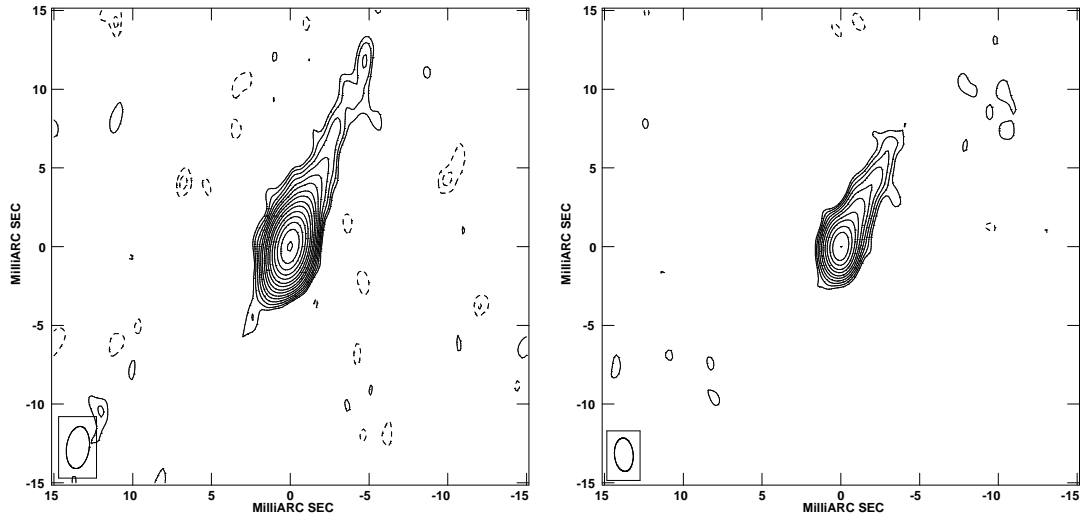

\centering 
\includegraphics[width=7cm, angle = -90]{June2009.ps} 
\includegraphics[width=7cm, angle = -90]{July2009.ps}
\caption{VLBA observations at 8.4~GHz in June and July 2009. Contours are at levels of $\pm(\sqrt{2})^n$ times the rms noise level, where n =  2, 3, 4, 5... a) Observed on 2009 June 28 (MJD 55\,010) with a peak brightness of 12.4~mJy/bm, total flux of 16.8~mJy and rms noise of $45 \mu$Jy. b) Observed on 2009 July 14 (MJD~55\,026) with a peak brightness of 10.9~mJy/bm, total flux of 18.8~mJy and rms noise of $123 \mu$Jy.} 
\label{fig:VLBA2009} 
\end{figure}

The first notable activity in 2009 began in June (around MJD~54\,985), with the X-ray emission rising up to 48~c/s (640~mCrab) in the ASM band and radio emission flaring up to $\sim30$~mJy; however, the hardness ratio remained above 1.2 (for 5-12~keV/3-5~keV) and Cygnus X-1 returned to the low/hard state, hence we observed a possible `failed' state transition~\cite{2003A&A...407.1039P}. Two VLBA observations were taken during the event. The first VLBA observation (Fig.~\ref{fig:VLBA2009}a) on MJD~55\,010 (at 8.4~GHz) showed just a compact jet with a peak brightness of $\sim12$~mJy bm$^{-1}$. The source was resolved, with the brightness distribution extending to both northwest and southeast of the peak. The second epoch on MJD~55\,026 (part of a separate project) was taken at both 1.4 and 8.4~GHz with peak brightnesses of 7.4 and 10.9 mJy bm$^{-1}$ respectively. The 1.4~GHz image was too scatter-broadened for us to detect any large scale structure, while the 8.4~GHz image (Fig.~\ref{fig:VLBA2009}b) detected a resolved one-sided jet, with extension only to the northwest of the peak. The X-ray emission then returned to its usual low/hard state.

\vspace{-0.3cm} 

\subsection*{Observations of summer 2010}

\begin{figure}[h]
\centering 
\includegraphics[width=7cm]{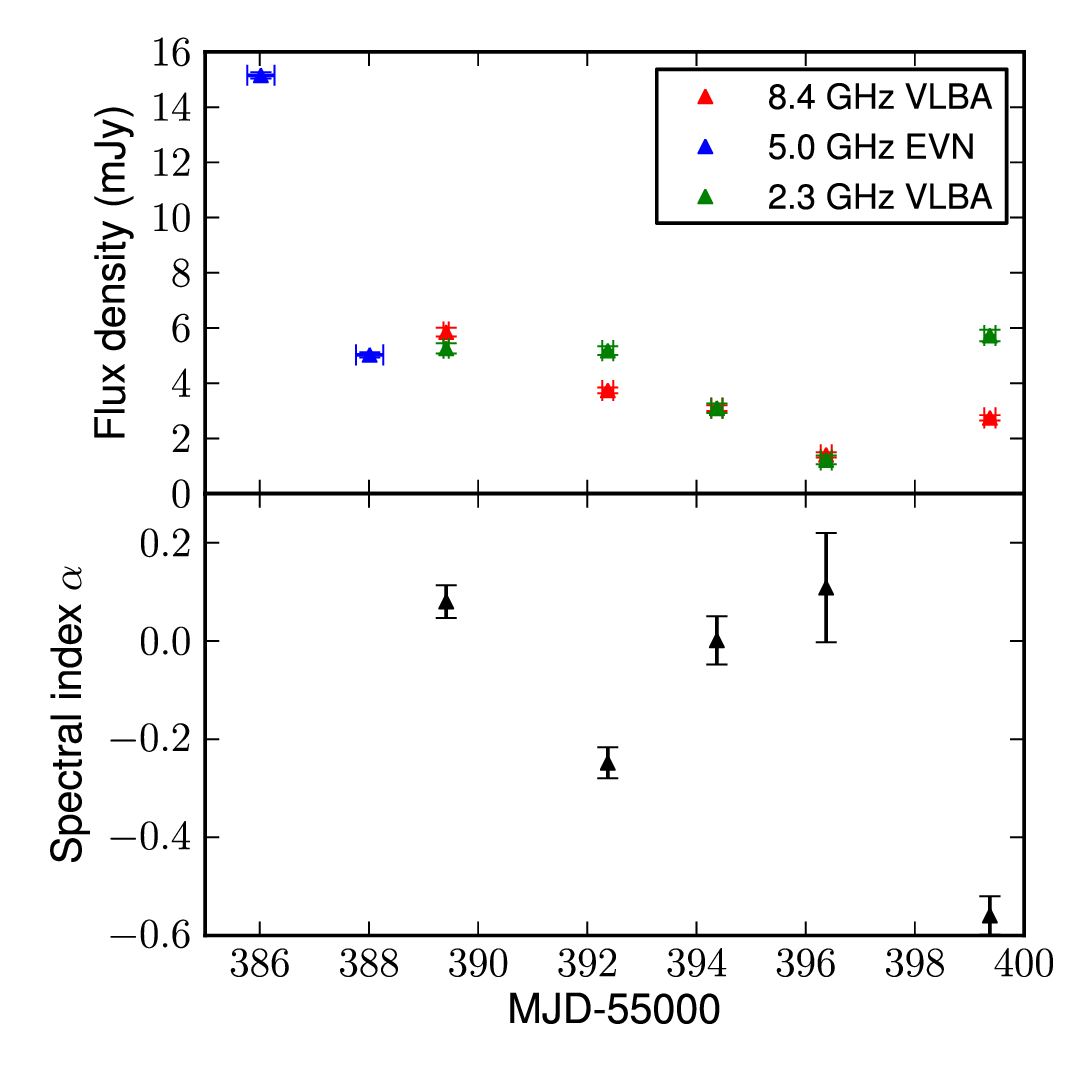} 
\includegraphics[width=7cm]{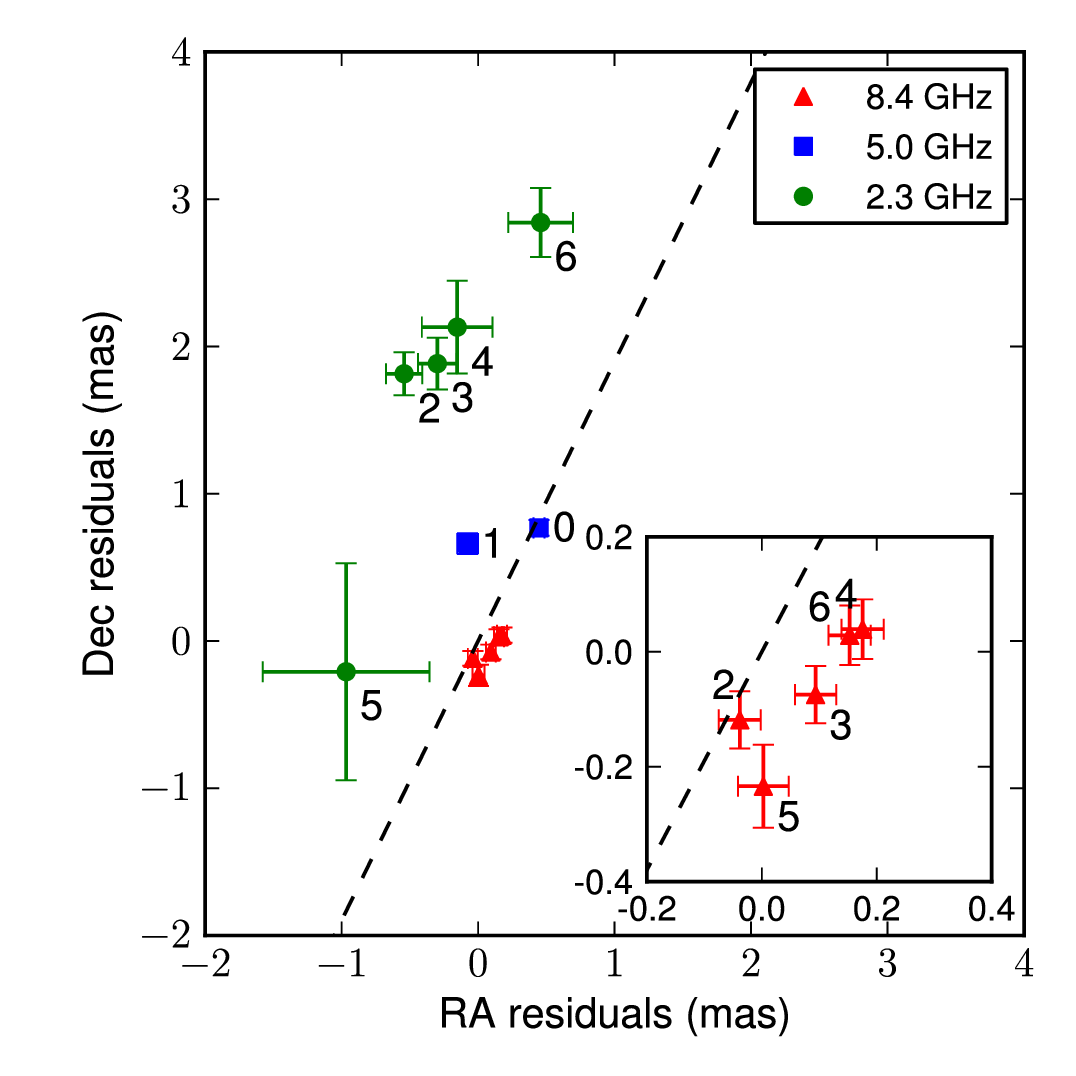} 
\caption{a) VLBI light curve and spectral index of Cygnus X-1 during the soft state in July 2010. b) Astrometric position of Cygnus X-1 during the soft state in July 2010 (note the positions are systematically offset as a function of frequency due to the phase calibration.  Inset shows a zoom-in of the 8.4\,GHz positions.  Numbers indicate the ordering of the 2010 epochs.} 
\label{fig:VLBI_results} 
\end{figure}

The {\it RXTE} ASM detected an increase in X-ray flux, starting around 2010 June 23 (MJD~55\,370) and increasing to a brightness of 52 c/s (690~mCrab) by 2010~July~1 (MJD~55\,377.8) \cite{2010ATel.2714....1R}. \linebreak {\it MAXI}/GSC~\cite{2010ATel.2711....1N} and {\it Fermi}/GBM~\cite{2010ATel.2721....1W} also detected increased activity, confirming the X-ray flaring event. Interestingly, the rise phase of the X-ray flare was not smooth, showing a short dip in X-ray count rate on 2010~June~29~(MJD~55\,375.9).  All VLBI observations in 2010 July detected Cygnus X-1 as an unresolved variable source. The two initial e-EVN observations detected a fast quenching of the radio emission from 16 to 5~mJy~bm$^{-1}$ over the $\sim48$ hours between epochs.  The source flux density remained $<6$\,mJy for all subsequent VLBA observations, although both the flux density and the spectral index between 2.3 and 8.4\,GHz were variable, as shown in Fig.~\ref{fig:VLBI_results}a.

Although the source was unresolved down to the beam size in all VLBI observations, we fitted the position of the centroid of the (unresolved) emission at each epoch.  We subtracted the estimated contributions of proper motion, parallax and orbital motion to the position at each epoch, and the residuals are shown in Fig.~\ref{fig:VLBI_results}b.  At both 8.4 and 2.3~GHz, the residuals are scattered along an axis which is aligned with the known position angle of the hard state jet (shown as the dashed black line).  The positional shift between different frequencies is an artifact caused by a frequency-dependent core shift in the phase reference source.

\section{Discussion and Conclusions}

There are stark structural differences in the jet between the hard state, the failed soft-state transition (during 2009) and the full transition to the soft-state (during 2010).  These structural changes show the value of high-resolution VLBI monitoring of the source during transitional states.  Integrated flux density monitoring with, for instance, the EVLA or WSRT, cannot provide these insights into the evolution of the radio jets.

The observation during the failed soft-state transition shows a similar morphology to the hard state jet, but at a higher luminosity, possibly suggesting that the outflow rate and/or velocity has increased (although Doppler `de-boosting' might be expected if the source is significantly relativistic). Such an increase in jet power ($Q_{\rm{jet}}$) would therefore associate with the slight increase in soft X-ray counts and hence mass-accretion rate ($\dot{m}$).  While it is tempting to interpret the southeastern extension as a counterjet in Fig~\ref{fig:VLBA2009}a, the peak emission in the hard state could come from the surface of optical depth unity, which should be some distance downstream from the central binary itself; thus, a more likely explanation for the extension would be a change in the optical depth and hence brightness distribution in the inner regions of the approaching jet.

Lastly, the absence of a jet on a size scale greater than a few mas during the full transition to the soft state is consistent with the model that the outflow in radio emitting XRBs is quenched below some arbitrary X-ray spectral hardness value (i.e. after crossing the `jet-line' in the HID), but with two major exceptions:

\begin{itemize}

\item It appears that Cygnus X-1 did not produce any `bright' discrete ejections, despite the source clearly transitioning from the low/hard state to the high/soft state.  However, one must consider that we did not have continuous coverage over the state transition, although missing the ejection is unlikely as the start of the state transition is well sampled and covers the predicted period of ejection (i.e. the `jet-line'). Therefore this unexpected result demonstrates that not all black hole candidates necessarily produce strong discrete ejecta during a state transition, which are thought to arise, for instance, from internal shocks forming in the jet due to an increase in the mass outflow rate and ejection velocity~\cite{2004MNRAS.355.1105F}.

\item The radio emission is not completely quenched after crossing the `jet-line'. We suggest that the alignment of the centroid positions along the jet axis is evidence for a jet in a soft state.  The scatter along the jet axis may arise from variations in parameters such as mass outflow rate or optical depth of either the jet itself or the surrounding stellar wind.  The non-detection of ballistically-moving discrete ejecta suggests that \textit{a weak compact jet with a variable spectral index is still present in the soft state of Cygnus~X-1}.

\end{itemize}


\section{Acknowledgements}

AR thanks ESO for funding a fellowship during this work. e-VLBI developments in Europe are supported by NEXPReS, an Integrated Infrastructure Initiative (I3), funded under the European Union Seventh Framework Programme (FP7/2007-2013) under grant agreement RI-261525. The European VLBI Network~\footnote{http://www.evlbi.org} is a joint facility of European, Chinese, South African and other radio astronomy institutes funded by their national research councils.  The National Radio Astronomy Observatory is a facility of the National Science Foundation operated under cooperative agreement by Associated Universities, Inc. The AMI is operated by the University of Cambridge and supported by STFC. The X-ray data was provided by the ASM/\textit{RXTE} teams at MIT and at the \textit{RXTE} SOF and GOF at NASA's GSFC and also thanks to the Swift team for the XRT observations.

\end{document}